# In-situ observation the interface undercooling of freezing colloidal suspensions with differential visualization method

Jiaxue You[1], Lilin Wang[2], Zhijun Wang[1*], Junjie Li[1], Jincheng Wang[1], Xin Lin[1], and Weidong Huang[1]

1-State Key Laboratory of Solidification Processing, Northwestern Polytechnical University, Xi'an 710072, P. R. China

2-School of Materials Science and Engineering, Xi'an University of Technology, Xi'an 710048, P. R. China

**Abstract:** Interface undercooling is one of the most significant parameters in the solidification of colloidal suspensions. However, quantitative measurement of interface undercooling of colloidal suspensions is still a challenge. Here, a new experimental facility and gauging method are designed to directly reveal the interface undercooling on both static and dynamic cases. The interface undercooling is visualized through the discrepancy of solid/liquid interface positions between the suspensions and its solvent in a thermal gradient apparatus. The resolutions of the experimental facility and gauging method are proved to be 0.01K. The high precision of the method comes from the principle of converting temperature measurement into distance measurement in the thermal gradient platform. Moreover, both static and dynamic interface undercooling can be quantitatively measured.

**Key words:** colloidal suspensions, directional solidification, interface undercooling, differential visualization method

## 1. Introduction

Solidification of colloidal suspensions is receiving increasing interest in a variety of scientific and industrial fields, such as cryobiology[1], tissue engineering[2], the fabrication of bio-inspired porous materials and composites[3, 4], thermal energy storage[5] and soil remediation[6]. In all these fields, controlling the solidification process of colloidal suspensions is a decisive factor to achieve desired results in terms of distributions of particles.[7]

*Corresponding author. Tel.:86-29-88460650; fax: 86-29-88491484
E-mail address: zhjwang@nwpu.edu.cn (Zhijun Wang)

In order to modulate such process, the key scientific issues therein are supposed to be understood clearly, i.e. the interactions between particles and advancing liquid/solid interface. For many years, the engulfment of an isolated single particle into the solid is considered to comprehend the basic physics behind the phenomenon, named isolated single particle models.[8-11] However, since colloidal suspensions consist of a great number of particles, the most common phenomena include a layer of particles building up against the interface.[12-15] All of the isolated single particle models failed in describing the interactions between the interface and the concentrated layer. Instead, the continuum-based thermodynamic model was proposed by Peppin et al.[16] in describing the solidification process of hard-sphere colloidal suspensions to simplify the complex interactions. Peppin et al.[17] discovered that the constitutional supercooling phenomenon in the system of colloidal suspensions is extremely similar to what has been observed for the solidification of binary alloys.

In recent years, interface undercooling in freezing colloidal suspensions is supposed to be of great importance on microscopic pattern formation.[16-19] For example, particulate constitutional supercooling caused by concentration gradients in colloidal system can trigger a thermodynamic instability that leads to morphological transitions in the pattern of segregated ice.[17] Moreover, these thermodynamic models were extended to cohesive particle systems, giving conditions for the formation of ice lenses.[20] The confusing characterization, that the existence of instability and meta-stability domains in cellular solidification of colloidal suspensions, attracts Deville et al.[21] attention and they found that these metastable and unstable crystal morphologies are essentially controlled by interface undercooling. Therefore interface undercooling, a crucial physical parameter, is of significant importance on expounding some puzzling phenomena in the colloidal system as mentioned above.

In the aspect of experiments, thermodynamic undercooling of colloidal suspensions is usually surveyed by differential scanning calorimetry (DSC) which measures the freezing point of the soil–water system through melting.[22] Although DSC can be utilized to quantitatively measure melting or freezing points, some drawbacks exist when it is applied to the system of colloidal suspensions. Firstly, the test precision is constrained because the temperature resolution of DSC usually can reach only 0.1K. Furthermore, the melting or freezing points are obtained from extrapolation, which also can bring some uncertainty. Secondly, the DSC method can only inspect static equilibrium melting point and fails to detect dynamic interface undercooling which is closely connected with microscopic morphology evolution and interfacial stability.[23] Thirdly, there are discrepancies between the theoretical prediction and the experimental data from DSC test.[17] The measured value of constitutional supercooling from DSC for volume fraction $\phi=0.5$ and a bentonite particle of hydrodynamic radius R =0.5 μm is almost 5K[17] whereas the theoretical prediction is about $6.09 \times 10^{-6}$ K[16] under the same conditions. So, a quantitative new method is eagerly needed to determine the interface undercooling of colloidal suspensions and further verify or settle the argument.

To explore this issue, here we propose a new gauging method with high resolution, by visualizing the interface undercooling through the discrepancy of solid/liquid interface positions between the suspensions and its solvent in a thermal gradient apparatus, named differential visualization (DV) method. The resolutions of the thermal gradient apparatus and the DV gauging method are analyzed. The high precision of the method comes from the principle of converting temperature measurement into distance measurement in the thermal gradient platform. Utilizing the DV method, the dynamic interface temperature of Polystyrene microspheres (PS) colloidal suspensions is quantitatively measured.

## 2. Experimental apparatus and gauging method

### 2.1 Experimental apparatus

As shown in Fig.1, the experimental platform we adopted is known as a horizontal Bridgeman directional freezing stage. This platform can avoid the convection interference caused by gravity. Furthermore, it aims at producing a constant and uniform thermal gradient along which samples of colloidal suspensions are pushed mechanically at a constant pulling speed. Similar experimental facility is utilized to simulate segregated ice formation during the unidirectional solidification of a colloidal suspension (bentonite).[17] However, it is not elaborate enough to research some significant details of interface undercooling due to its large size of the Hele-Shaw cell (380×120×5 mm) and the large gap between the heating and cooling zones (60 mm).[24] This large gap can't guarantee a high precision for thermal gradient which is of great importance on the measurement of interface undercooling. Here we redesigned the experimental apparatus to overcome these drawbacks.

In our experimental apparatus, the thermal gradient is produced by two heating and cooling zones separated by a gap of 5 mm. The temperatures of both heating and cooling zones are provided by ethanol thermostat whose temperature can change from -20.0 ℃ to 50.0 ℃ discretionarily. The temperature of the thermostat was set by a temperature controller. Sample translation across the thermal gradient is provided by a servo-drives motor supplemented with a linear ball-screw drive. Observation of the growth front is achieved through an optical microscope stage with a charge-coupled device (CCD) camera.

In both the cooling and heating zones, samples are sandwiched by temperature-controlled surfaces in close contact with them. Both heaters and coolers are made of copper blocks (120×100×10 mm) and ethanol circulation stages. The part of temperature gradient is thermally insulated by asbestos plates in order to minimize lateral thermal losses and thus obtain linear temperature gradient.

Sample translation is provided by a linear ball-screw driven stage. The rotation of the ball screw is actuated by a servo-drives motor through a decelerator. The whole pulling equipment is fixed on a big cast-iron experiment table to eliminate exotic mechanical vibration. The screw has a 4-mm pitch, linking it rigidly to the motor axis. The motor is fixed on micro-positioners in order to align the screw on the track accurately.

Crystalline interfaces were observed according to the usual method based on the evidence of optical aberration induced on an incident parallel light beam by refraction at the growth front. In order to minimize growth perturbations, the microscope stage can be moved up and down smoothly for focusing accurately. Images were recorded *via* a charge-coupled device (CCD) camera with 2580×1944 sensitive elements on a time-lapse video recorder and further analyzed by the software image processing.

**2.2 Gauging method**

Thermodynamic undercooling of colloidal suspensions is usually measured by DSC[22] or thermo-couple[23] in previous research. Here we introduce a new gauging method and compare it with DSC and thermo-couple methods.

According to the phase diagram of colloidal suspensions[16, 17], the equilibrium freezing point of the suspensions is lower than that of its solvent. Thus, under a certain temperature gradient, interface position of the suspensions is expected to be closer to the cooling zone and then an interface gap between the suspensions and its solvent will appear. For this reason, the sample cell of colloidal suspensions (sample 1) is placed side by side with a cell of its solvent (sample 2) under an identical thermal gradient, shown as Fig.2A. Interface gaps between two samples are recorded and measured through snapshot. The separation distance between melting front of the solvent (red dash line in Sample 2) and interface position of suspensions (blue dash line in Sample 1) indicates the interface undercooling of colloidal suspensions as visually shown in Fig.2. Through the difference of pixels ($\Delta S$) between these two lines on the photograph combined with image scale (M) and temperature gradient (G), the interface undercooling ($\Delta T$) of colloidal suspensions is calculated as $\Delta T = S \times G$ ($S = \Delta S \times M$ is the real distance of the interface gap). Note that, the interface undercooling ($\Delta T$) will turn into interface temperature ($T_i$), if the solvent in Sample 2 is replaced by ultrapure water. Fig.2A is the schematic diagram of this gauging method, while the real interface gap during the dynamic cellular growth is exhibited in Fig.2B. On the one hand, compared with DSC, this method can investigate dynamic interface temperature and undercooling. On the other hand, compared with interface temperature measurement utilizing thermo-couple directly, this method is more convenient and efficient.

# 3. Evaluations of experimental facility and gauging method

## 3.1 Accuracy of experimental facility

For the accuracy of pulling velocity of our experimental facility, it is measured in real time using screw thread micrometer to record distance (D) and second chronograph to record time (T). The authentic pulling speed (V) is calculated as V=D/T. After several measurements of the identical pulling velocity, the average pulling speed is obtained so as to counteract the statistical error.

Precautions were taken to improve uniformity of the thermal gradient and reduce its fluctuations. Compared to the width of the thermal devices (100 mm), a much smaller width of samples (2 mm) was chosen to suppress edge effects of the thermal devices. In addition, these lateral sides of the thermal devices were thermally insulated by asbestos plates so as to minimize lateral thermal losses. On the other hand, in order to reduce radiative and convective perturbations, the thermal blocks were covered with heat-insulating shield and the gap between hot and cold copper blocks were covered with double-glazed windows, which had both excellent thermal insulation and optical microscopic observation. Also, the whole experimental apparatus is placed in the environment of constant low temperature (around 16 degrees Celsius). Finally, sufficiently large copper blocks (120×100×10 mm) were used in each thermal device so as to filter external perturbations and reduce their effects on the temperature field of the samples. Direct observation *via* CCD has also confirmed the absence of measurable disturbances, either time dependent or permanent, on steady interfaces on both static and dynamic cases, as shown in the Movies S1 and S2 (Supporting Information)[25].

However, on the dynamic case, the thermal field constructed by conduction is also affected by the dynamic sample translation, which will induce a nonlinear displacement of the isotherms towards the cold boundary and thus possibly a non-uniform thermal gradient. These effects increase with pulling velocity, resulting in increased drifts of both the melting isotherm and thermal gradient G on the interface. The dynamical interface drift in the dynamic transient stage is caused by three different factors[26]: 1. An isotherm shift due to pulling that is deduced from the comparison of the liquidus isotherms positions at rest and during pulling for $\Delta H = 0$; this shift of isotherms can be seen as the "instrumental recoil", which is due to the evolution of thermal exchanges induced by pulling; 2. The solutal recoil that corresponds to the interface temperature change from the liquidus to the solidus (taken during pulling for $\Delta H = 0$); 3. An isotherm shift due to latent heat release measured from the comparison of the solidus isotherms positions during pulling for $\Delta H = 0$ and the normal value of $\Delta H$. Type 2 is an expected physical phenomenon while type 1 and type 3 need to be eliminated. The design of side-by-side samples in Fig.2 can almost surmount these two thermal lag effects and obtain the interface undercooling of colloidal suspensions more precisely. The instrumental recoil (type 1) can be counteracted completely by the comparison with the solvent. The isotherm shift (type 3) can be partly eliminated by the comparison with the solvent since the solvent freezing also releases latent heat. In addition, the pulling velocity in the experiments is very low (less than 16μm/s) which can only cause a feeble influence on thermal gradient. In the following, the local thermal gradient around the interface under varying pulling speeds will be investigated to ascertain the adverse effect caused by sample translation.

Figure 3 presents the temperature gradients on the samples both in static and dynamic process. The temperature was measured by thermocouple. Before use, the thermocouple was calibrated by a standard platinum resistance thermometer with an uncertainty of ±1 mK. Static temperature gradient was measured based on seven discrete points in the samples, while the dynamic thermal gradients are investigated through Yokogawa LR 4110 temperature recorder (Japan, model: 371123). What's worth mentioning, only one thermocouple was employed to measure the temperature profile because the systematic measurement error of the thermocouple can be removed when the static temperature gradient was calculated. Figure 3 shows that under all pulling speeds, the thermal gradients have extremely excellent linearity (R=0.99876 for V=16μm/s, R=0.99778 for V=7.63μm/s, R=0.99209 for V=3.486μm/s and R=0.99981 for V=0). Moreover, the local temperature gradients change little with different pulling speeds since G=7.33K/cm for V=16μm/s, G=7.28K/cm for V=7.63μm/s, G=7.03K/cm for V=3.486μm/s, and G=7.01K/cm for V=0. There is only a slight increase of thermal gradient for large pulling speed probably because the heat from hot plate is taken to heating zone *via* sample translation. It warms the heating zone and the cooling zone keeps constant due to cold plate. Thus the thermal gradient is enlarged a little. Nevertheless, the maximum relative error of dynamic thermal gradient caused by pulling velocity is less than 4.6% and this delicate difference can be neglected when the pulling speed is small (less than 16μm/s).

The error of the calculated interface undercooling $\Delta T$ (=$S \times G$) derives from the real distance of the interface gap S and thermal gradient G. The statistical error of G ($\delta G$) from multi-measurements is less than 0.1K/cm. Image resolution of CCD can reach to 1 pixel pitch (=1.08 μm) and the interfacial thickness on the photograph is about three pixels. Thus the error of S ($\delta S$) is about 3.24 μm. Therefore the error of the calculated interface undercooling ($\delta \Delta T = \delta S \times G + \delta G \times S$) is less than $5.1 \times 10^{-3}$ K. The resolution can reach to 0.01K since the error comes from the thousandth.

**3.2 Resolution verification of gauging method**

Next dilute sodium chloride aqueous solution is utilized to examine the resolutions of the gauging method. Ultrapure water (provided by deionizer when the resistance of water comes to 18.25 MΩ) and sodium chloride (AR, 99.5%) are prepared to compound sodium chloride aqueous solution of 0.158%, 0.105%, 0.079% and 0.0395% weight fraction. Using the DV gauging method, the constitutional supercoolings of the prepared dilute sodium chloride aqueous solutions are measured several times under the same thermal gradient. The error bars in Fig.4 show that the statistical errors under an identical temperature gradient can be ignored. The comparison between the measurements and the phase diagram of NaCl-$H_2O$ system[27] is shown in Fig.4. The measurements are consistent with the phase diagram of dilute NaCl-$H_2O$ system. The theoretical value of constitutional supercooling for 0.0395 wt% NaCl-$H_2O$ is 0.0718K and the measurement from this gauging method is 0.0706K. The measured result is shown in Movie S1 (Supporting Information)[25]. Therefore the resolution can reach to percentile (0.01 K). We also use different thermal gradients to measure the undercooling of identical dilute sodium chloride aqueous solution as shown in the inset of Fig.4. It exhibits that the measurements are robust for different temperature gradients and the errors of undercooling from different thermal gradients are very small (less than 0.005K).

It should be noted that the resolution is also determined by the thermal gradient. Within a smaller thermal gradient, the pixels in the images from CCD will represent a smaller temperature variation. However, a small thermal gradient will be disturbed easily by the ambient thermal fluctuation. Therefore, the thermal gradient should be in a finite range.

## 4. Interface temperature of PS colloidal suspensions

We choose Polystyrene microspheres (PS) aqueous suspensions with a diameter of 100±1 nm (Bangs Lab) in the experimental investigations. One of the advantages of these suspensions is the weak sedimentation effect. Before experiment, the suspension is homogenized through ultrasonic cleaners for 30 minutes to further avoid the weak settling effect. 100 μm thick rectangular glass capillary (VitroCom, model: RT5010, 100×2 ×0.1 mm) was adopted as sample cell. The PS suspension was introduced by capillarity before sample airtightness was achieved by further gluing. The prepared sample is placed on a glass slide (1 mm thick) and glued with araldite. Here we mainly considered the dilute case, and samples with volume fraction $\phi_0$=10% are prepared. The sample of ultrapure water is placed side by side with that of PS suspensions. So the measurement is the interface temperature as mentioned in the part of 'Gauging method'. The prepared sample cell is homogenized in the experimental setup for 30 minutes before it freezes.

Based on the excellent jarless thermal gradients, the dynamic interface temperature of PS colloidal suspensions on varying pulling speeds is detected by the DV gauging method, as shown in Fig.5. Although the thermal gradient G=8.72K/cm is for V=0, the thermal gradients for other pulling speeds are derived on the basis of variation trend of Fig.3 and further utilized to calculate the dynamic interface temperature. According to the inset of Fig.5, the interface position of colloidal suspensions is always lagging behind the melting front of ultrapure water under different pulling velocities. With the increase of pulling speed, the gap between the solidifying interfaces of ultrapure water (red dash line) and suspensions (blue dash line) decreases and hence the interface temperature rises close to 0 degree Celsius. Moreover, under a relative high pulling speed, seaweed dendrite morphology appears as shown in inset A of Fig.5. With the decrease of pulling velocity, seaweed dendrite-cellular transition happens. From the measurements, dynamic interface temperature can be observed in-situ and the resolution of the DV gauging method can reach to 0.04K. Furthermore, microstructure evolution is visualized during the steady-state growth of colloidal suspensions, as shown in Movie S2 (Supporting Information)[25].

## 5. Conclusion

Interface undercooling is one of the most significant physical parameters since particulate constitutional supercooling phenomenon is discovered in the system of colloidal suspensions in recent years. It is achieved that quantitative measurement on interface undercooling of colloidal suspensions *via* a new experimental facility. Both static and dynamic interface undercooling are visualized through a new gauging method, i.e. the comparison of interface positions between the suspensions and the solvent in a thermal gradient apparatus. The resolution of the experimental apparatus and the gauging method are analyzed and measured. The part of temperature gradient which is of great importance on the whole facility is with high linearity on both static and dynamic cases. The sensibility of the gauging method has been confirmed utilizing dilute sodium chloride aqueous solution and its phase diagram. The high precision of the method comes from the principle of converting temperature measurement into distance measurement in a thermal gradient platform. The dynamic interface temperature in colloidal system is measured under varying pulling velocities.

With this carefully choreographed measuring method, we can survey easily the equilibrium freezing point of colloidal suspensions and their dynamic interface undercooling and temperature during solidification. In addition, microstructure evolution of freezing colloidal suspensions can be observed in-situ during the steady-state growth.

**Acknowledgements**

This research has been supported by National Basic Research Program of China (Grants No.2011CB610401), Nature Science Foundation of China (Grant No. 51371151), Free Research Fund of State Key Laboratory of Solidification Processing (100-QP-2014), the Fund of State Key Laboratory of Solidification Processing in NWPU (13-BZ-2014) and the Fundamental Research Funds for the Central Universities (3102015ZY020).

**List of figures**

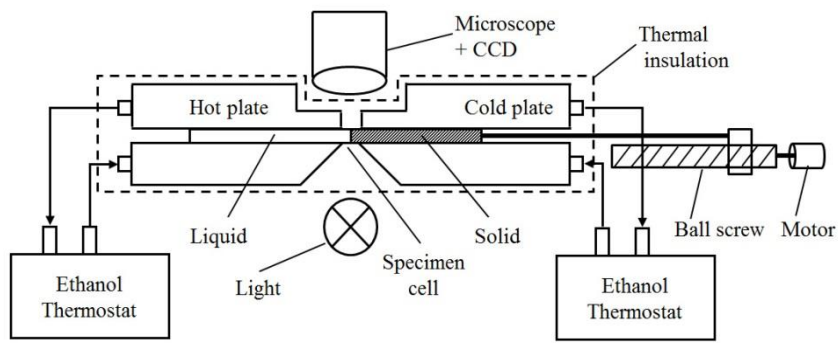

Fig. 1 Schematic diagram of horizontal directional freezing stage.

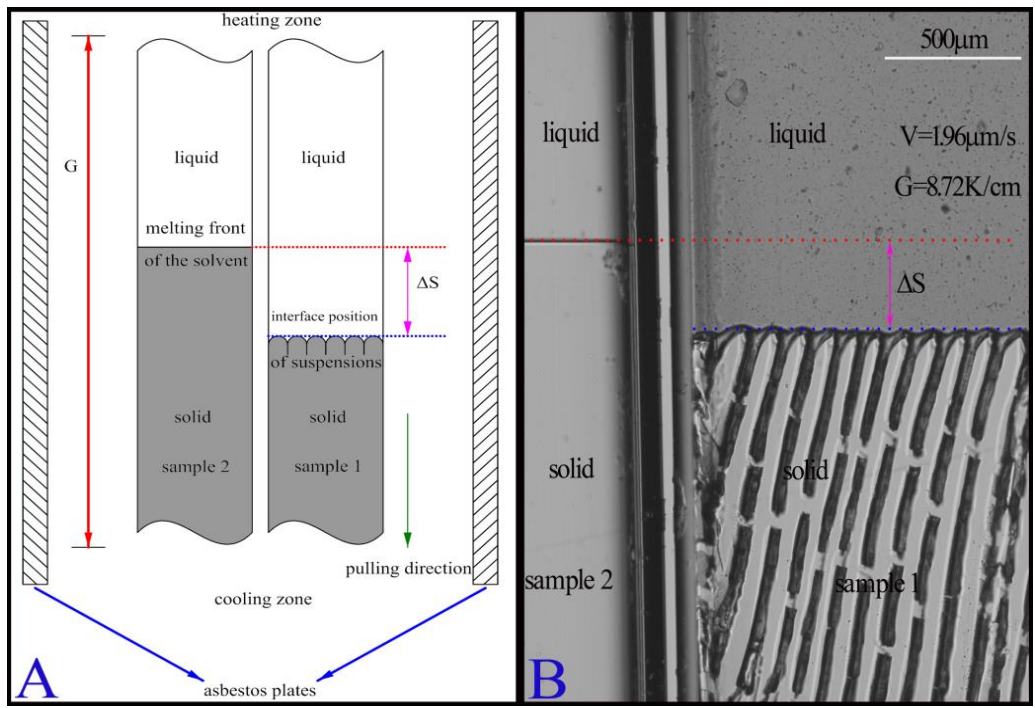

Fig.2 Schematic (A) and matter (B) gap between the sample cells in the gauging method. The cellular morphology appears on condition of G=8.72K/cm, pulling velocity V=1.96μm/s, initial volume fraction $\phi_0$=10% and particle diameter d=100nm for Polystyrene microspheres (PS) aqueous suspensions.

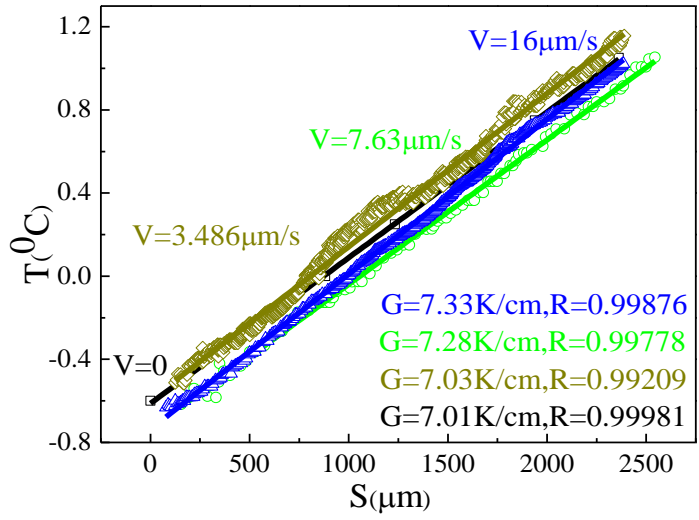

Fig.3 actual measurement of local thermal gradient on the interface under varying pulling speeds.

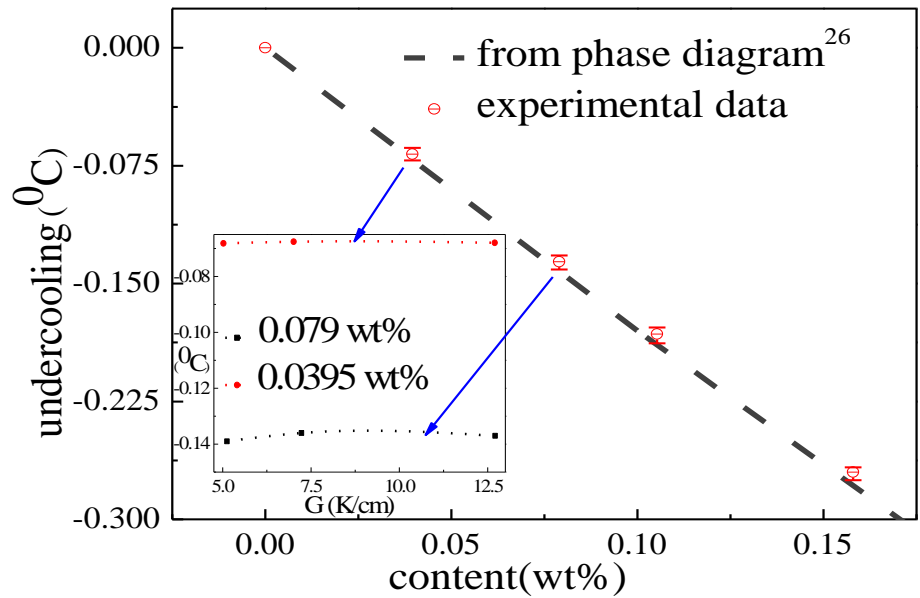

Fig.4 static undercooling of dilute NaCl-H$_2$O solution under identical thermal gradient (G=7.01K/cm) compared with its phase diagram. The inset is the undercoolings of 0.079 and 0.0395 wt% sodium chloride aqueous solution under different thermal gradients.

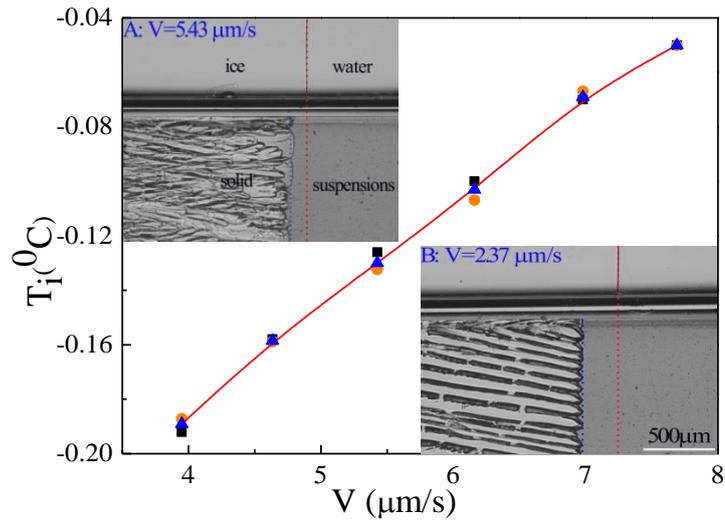

Fig.5 dynamic interface temperature of colloidal suspensions under varying pulling velocities with temperature gradient G=8.72K/cm for V=0, initial volume fraction $\phi_0$=10% and particle diameter d=100nm. The different colors refer to separate experiments. The insets A and B are schematic matter comparison under different pulling speeds.

# In-situ observation the interface undercooling of freezing colloidal suspensions with differential visualization method

Jiaxue You[1], Lilin Wang[2], Zhijun Wang[1*], Junjie Li[1], Jincheng Wang[1], Xin Lin[1], and Weidong Huang[1]

1-State Key Laboratory of Solidification Processing, Northwestern Polytechnical University, Xi'an 710072, P. R. China

2-School of Materials Science and Engineering, Xi'an University of Technology, Xi'an 710048, P. R. China

## Supporting Information

## Movies S1 and S2

Movie S1: The measured static interface undercooling of 0.0395 wt% $NaCl-H_2O$ for thermal gradient G=5.02K/cm. The interfaces are steady. It shows that the theoretical value of constitutional supercooling for 0.0395 wt% $NaCl-H_2O$ (0.0718K) can be easily recognized through our DV gauging method.

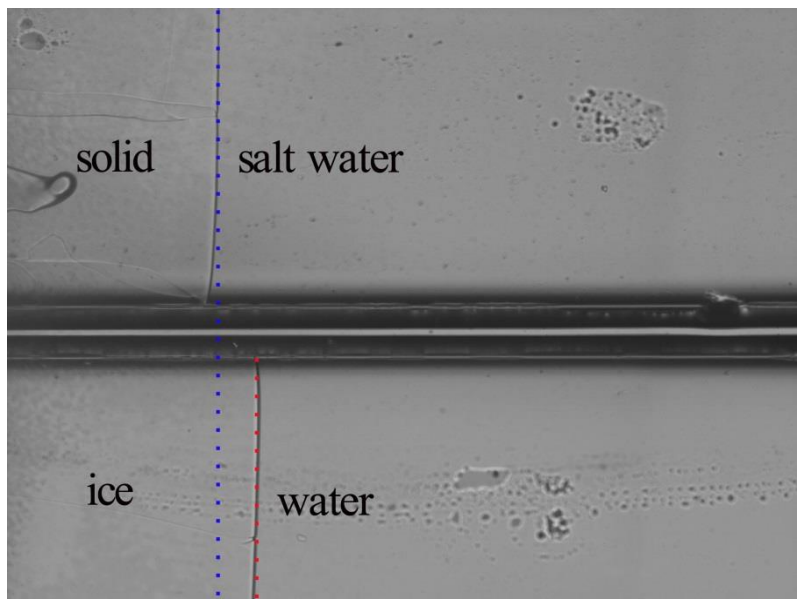

Movie S1 (Multimedia view)

*Corresponding author. Tel.:86-29-88460650; fax: 86-29-88491484
 E-mail address: zhjwang@nwpu.edu.cn (Zhijun Wang)

Movie S2: The dynamic steady-state growth of PS colloidal suspensions (the cellular morphology) and ultrapure water (planar interface). G=8.72K/cm, V=2.37μm/s, initial volume fraction $\phi_0$=10% and particle diameter d=100nm. The scale bar is the same as Movie S1's.

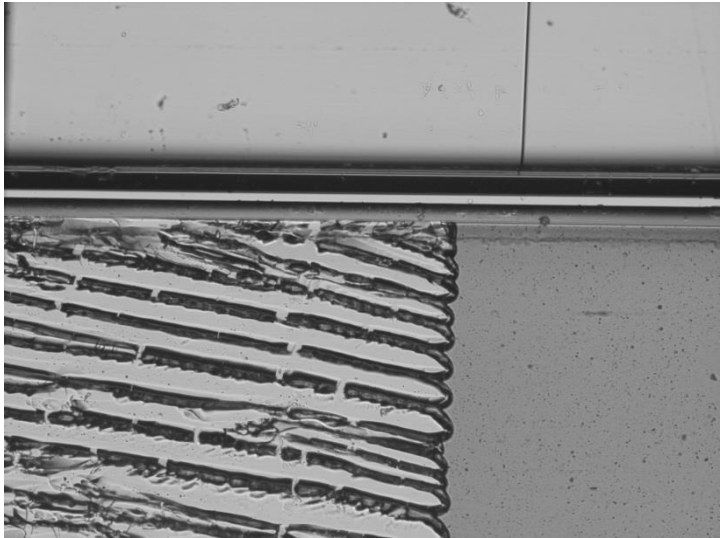

Movie S2 (Multimedia view)